Beyond pass or fail: an information-processing account of the Turing Test

Qiwei Zhao[1,2,3], Oriana Pansardi[1,#], James W. Strachan[1,4,#], Eugenio Scaliti[1,5,6,7,#], Guido Manzi[8], Giulia Borghini[1,9], Fabio Manzi[8], Alessandro Rufo[8], Frédérique de Vignemont[10], Michael S. A. Graziano[11], Stefano Panzeri[2], Cristina Becchio[1,*]

1. Department of Neurology, University Medical Center Hamburg-Eppendorf (UKE), Hamburg, Germany
2. Institute for Neural Information Processing, Center for Molecular Neurobiology (ZMNH), University Medical Center Hamburg-Eppendorf (UKE), Hamburg, Germany
3. School of Psychology, Shanghai University of Sport, Shanghai, China
4. Faculty of Governance, Economics, & Social Sciences, Université Mohamed VI Polytechnique, Rabat, Morocco
5. Department of Psychology, University of Turin, Turin, Italy
6. Department of Management “Valter Cantino”, University of Turin, Turin, Italy
7. Human Science and Technologies, University of Turin, Turin, Italy
8. Indeep Artificial Intelligence, Las Palmas de Gran Canaria, Spain
9. Child and Adolescent Neuropsychiatric Service (UONPIA), Fondazione IRCCS Ca' Granda Ospedale Maggiore Policlinico, Italy
10. Institut Jean Nicod, Département d’études cognitives, École normale supérieure, EHESS, PSL University, CNRS, Paris, France
11. Princeton Neuroscience Institute, Princeton University, Princeton, NJ, USA

# These authors contributed equally

* To whom correspondence may be addressed: c.becchio@uke.de

**Abstract**

The Turing Test is passed when a human interrogator cannot distinguish machine-generated responses from human ones. Yet behavioral indistinguishability can reflect two fundamentally different failures: a failure of encoding, in which machine and human output distributions overlap so closely that source information is effectively absent; or a failure of readout, in which source information is present but the interrogator fails to extract it. To dissociate these possibilities, we inverted the Turing Test, placing humans and AI systems in symmetrical interrogator roles. Using a quantitative information-processing framework, we separated *encoding*, the information contained in the text; *readout*, the extraction of that information by an interrogator; and *meta-readout*, the interrogator's explicit access to the features guiding its own judgments. Across two series of experiments involving more than 3800 human interrogators and two generations of GPT models (GPT-4 and GPT-5), we found a striking dissociation across processing levels and interrogator types. As AI-generated responses became more human-like from GPT-4 to GPT-5, source information became more difficult for human interrogators to extract. GPT-5 interrogators, by contrast, read out source-discriminating information more accurately than GPT-4 interrogators. This improvement, however, came with a specific limitation: GPT-5 explanations only partially aligned with the feature weightings that actually predicted their judgments. Human interrogators showed the complementary pattern, with modest readout alignment but comparatively higher meta-readout alignment. Together, these findings suggest a double dissociation between readout and meta-readout across human and GPT interrogators.

## Introduction

The original Turing Test [1] was designed as an imitation game: a human interrogator attempts to distinguish between a man and a woman on the basis of a text exchange. At some point during the game, the man is replaced by a machine. If the interrogator remains incapable of distinguishing the machine from the woman, then the machine is said to have passed the test [2].

Turing offered this test as a philosophical thought experiment rather than an experimental tool to probe machine intelligence [3], yet decades of research have adopted this framework or variants of it to benchmark AI systems [2, 4, 5]. The logic is straightforward: if a machine can produce responses that a human interrogator cannot reliably distinguish from those of a human, then the machine can be said to have passed the test.  Seen that way, the Turing Test becomes a test of (in-)distinguishability [6].

But what does behavioral indistinguishability from humans actually demonstrate?  This question has regained urgency following recent claims that chatbots such as OpenAI's ChatGPT have finally passed the Turing Test [3, 7]. The issue is not merely whether these systems actually did so, but what such a result would mean: if failing the test proves nothing [4], what does passing the test prove?

A central problem is that behavioral indistinguishability can arise for fundamentally different reasons. One possibility is a *failure of encoding* [8, 9, 10, 11]: the machine's output distribution matches the human output distribution, such that the information required to discriminate between the two sources is effectively absent. In this case, responses are drawn from indistinguishable distributions, and no observer—however skilled—could perform better than chance. This would constitute a strong result, implying that machine-generated responses are, in measurable terms, equivalent to human-generated ones.

A second possibility is a *failure of readout* [8, 9, 10]: the relevant source information is present but not effectively extracted or used by the human interrogator [12]. Here, indistinguishability reflects limitations in the observer rather than equivalence of the underlying representations. A different interrogator, for instance, an artificial system with more effective readout capabilities, might succeed in discriminating between human- and machine-generated responses. Under this interpretation, the outcome of the Turing Test reveals less about machine intelligence than about the constraints of the interrogator [3].

These two possibilities are conflated in standard implementations of the Turing Test. Here we propose an approach to distinguish between these possibilities by recasting the Turing Test as an information-processing problem of encoding and readout [10].

First, at the level of *encoding*, we ask how much information about the generating source (human vs. AI) is present in the responses and which linguistic features encode that information. Second, at the level of *readout*, we complement the classical Turing Test with an *Inverse Turing Test* [13], testing both human participants and artificial systems in the role of interrogators. This enables a direct comparison of human and artificial readout computations, revealing not only whether humans and artificial systems can discriminate between response sources, but how they weigh and combine linguistic features to do so. Third, at the level of *meta-readout*, we ask whether interrogators can accurately report the features that guided their discrimination choices. Effective readout does not entail accurate self-report: an interrogator may correctly weight source-discriminating features to infer the source of a response without being able to describe that weighting [14]. The meta-readout level tests this potential gap directly: whether interrogators, human or AI, have access to their own readout computations.

We applied this framework to two generations of GPT models, GPT-4 and GPT-5, and compared their performance with that of a large sample of human participants (N = 3824). Results across experiments revealed a double dissociation across processing levels and interrogator types. As AI-generated responses became more human-like from GPT-4 to GPT-5, source information became more difficult for human interrogators to extract. GPT-5 interrogators, by contrast, read out source-discriminating information more accurately than GPT-4 interrogators. This improvement, however, came with a specific limitation: GPT-5 explanations only partially aligned with the feature weightings that actually predicted their judgements. Human interrogators showed the complementary pattern, with modest readout alignment but comparatively higher meta-readout alignment. Together, these findings suggest a double dissociation between readout and meta-readout across human and GPT interrogators.

## Results

Human and AI responses were elicited by using the *Strange Stories* test [15], a theory-of-mind paradigm designed to probe complex mental-state reasoning through short narrative scenarios involving deception, irony, persuasion, or misunderstanding. Responses generated by human participants and artificial systems to the same stories were then paired and presented to interrogators in a two-alternative forced-choice (2AFC) design. On each trial, the interrogator, human or AI, viewed a story followed by two responses, one human-generated and one AI-generated, presented in two consecutive intervals. Interrogators were instructed to identify the interval containing either the human-generated response (Human task) or the AI-generated response (AI task) and then provided a free-text explanation of the features that informed their judgment. Response order and task were counterbalanced across trials. We conducted two series of studies. In the first, human- and GPT-4-generated responses were evaluated by both human and GPT-4 interrogators; in the second, human- and GPT-5-generated responses were evaluated by both human and GPT-5 interrogators.

### Behavioral (in-)distinguishability

As a first step to characterize the extent to which interrogators could reliably distinguish human-generated from AI-generated responses, we applied Signal Detection Theory (SDT) [16]. SDT separates performance into sensitivity (d'), the observer's ability to discriminate between human-generated and AI-generated responses, and criterion (c), reflecting systematic biases toward selecting one interval over the other. Fig. 1 shows the results for the two series of studies comparing human responses against GPT-4 and GPT-5 responses, respectively.

In the first series of studies, human interrogators discriminated above chance in both task conditions (AI task: d' = 1.36, 95% CI [1.18, 1.53], $p = 0.008$; Human task: d' = 1.10, 95% CI [0.93, 1.28], $p = 0.008$), with a small discrimination advantage for the AI task over the Human task ($p = 0.018$) and no systematic criterion bias (both $p \geq 0.58$; Fig. 1a). Neither discrimination sensitivity nor criterion varied systematically as a function of participants' self-reported use of chatbots (Fig. S2a).

GPT-4 interrogators, by contrast, showed a marked asymmetry across tasks. Sensitivity was above chance in the AI task (d' = 0.79, 95% CI [0.60, 0.99], $p = 0.008$) but systematically reversed in the Human task (d' = -1.11, 95% CI [-1.32, -0.89], $p = 0.008$), resulting in a large and significant difference in d' between tasks ($p = 0.004$). This pattern was accompanied by a pronounced negative criterion, particularly in the AI task (AI task: c = -0.715, 95% CI [-0.80, -0.62], $p = 0.008$; Human task: c = -0.37, 95% CI [-0.44, -0.29], $p = 0.008$; Fig. 1a), indicating a systematic tendency to report that the target appeared in the first interval. Across both tasks, sensitivity was lower for GPT-4 than for human interrogators (AI task: p = 0.004; Human task: p = 0.004).

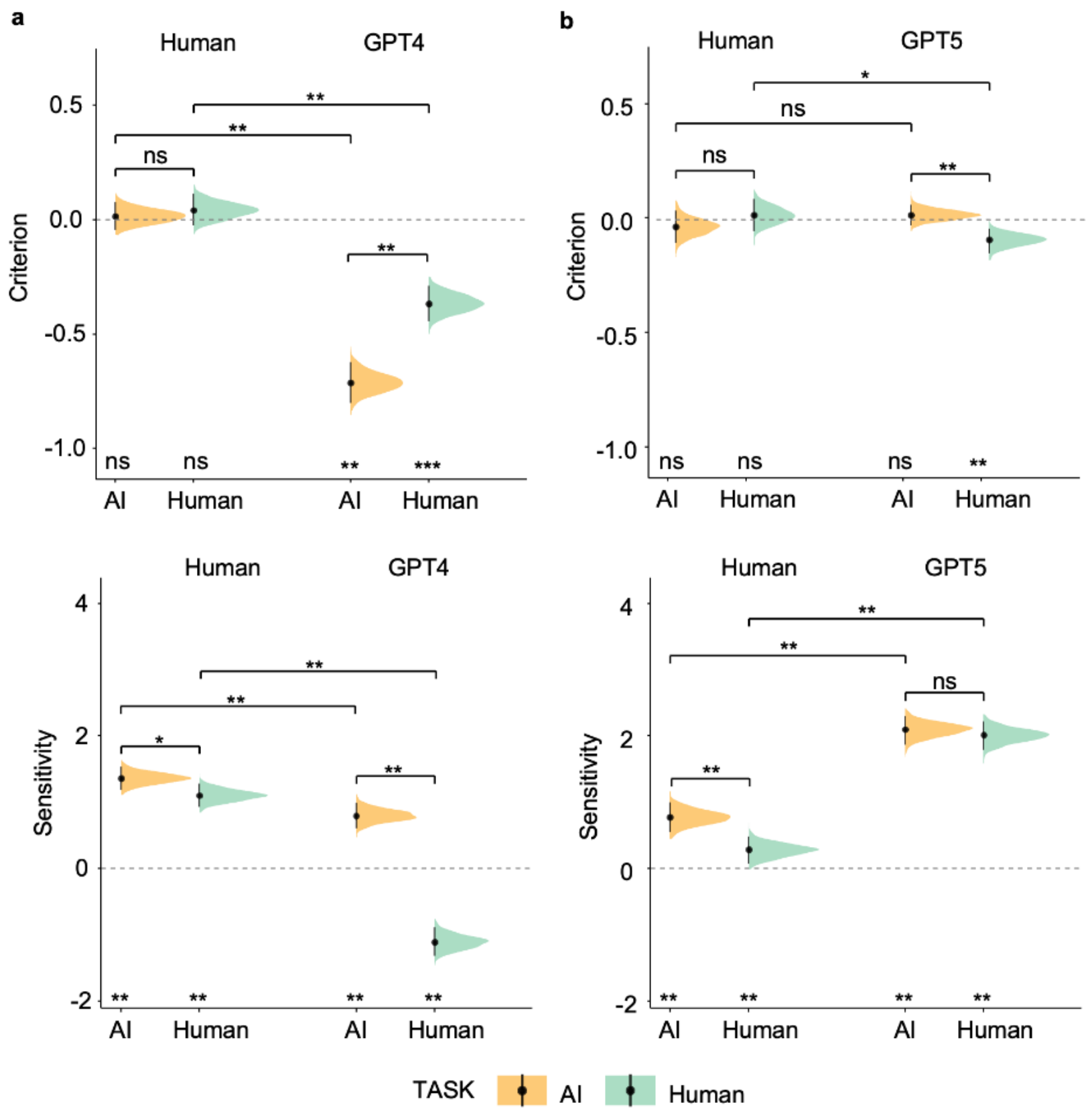

**Figure 1. Signal Detection Theory (SDT) analyses: Interrogators' performance in distinguishing human- from AI-generated responses.** Results are shown separately for **a)** the first series of studies, comparing Human and GPT-4 interrogators evaluating human- and GPT-4-generated responses, and **b)** the second series of studies, comparing Human and GPT-5 interrogators evaluating human- and GPT-5-generated responses. Column labels indicate the interrogator. X-axis labels indicate the task. Upper panels display criterion and lower panels display sensitivity estimates, each broken down by interrogator type (Human vs. AI) and task type (AI Task vs. Human Task). Points indicate posterior mean estimates; violin plots show bootstrap distributions. Asterisks below distributions denote significant deviations from zero; brackets with asterisks indicate significant between-condition differences ($p < .05$, ** $p < .01$, *** $p < .001$; ns = not significant).

In the second series of studies, the relative performance of human and AI interrogators reversed, with GPT-5 outperforming human interrogators across both tasks (AI task: p = 0.004; Human task: p = 0.004). As in the first series, human interrogators discriminated above chance between human- and GPT-5-generated responses in both tasks (AI task: d' = 0.77, 95% CI [0.54, 0.99], p = 0.008; Human task: d' = 0.27, 95% CI [0.07, 0.48], p = 0.008; Fig. 1b), with higher sensitivity in the AI task than in the Human task (p = 0.004). In the Human task, sensitivity did not vary systematically as a function of participants' self-reported use of chatbots; in the AI task, by contrast, sensitivity was higher among participants reporting more frequent chatbot use (Fig. S2a).

GPT-5 interrogators achieved high, symmetric discrimination across both task conditions (AI task: d' = 2.09, 95% CI [1.87, 2.29], p = 0.008; Human task: d' = 2.01, 95% CI [1.79, 2.22], p = 0.008), with no reliable task difference in sensitivity (p = 0.094; Fig. 1b). Criterion biases were also substantially reduced relative to GPT-4: GPT-5 showed no systematic criterion bias in the AI task (c = 0.02, 95% CI [-0.03, 0.07], p = 1.000) and only a small negative criterion in the Human task (c = -0.09, 95% CI [-0.15, -0.04], p = 0.008; Fig. 1b).

Taken together, these results establish that interrogators, both human and AI, could discriminate above chance between human- and AI-generated responses, but with markedly different sensitivity across Large Language Model (LLM) generations.

**Encoding**

Having established that both human and AI interrogators could behaviorally distinguish human- from AI-generated responses, the next step was to quantify the extent to which responses themselves differed at the level of encoding, that is, how much information about the generating source (human vs. AI) was present in the responses themselves and which linguistic features carried that information. To do so, we trained logistic encoding models to predict the generating source from differences in linguistic features including length, detail, typos, redundancy, punctuation, vocabulary complexity, and conversational tone (Fig. 2a; see Methods for the full feature set). Encoding accuracy was quantified as the balanced classification accuracy of these models. Higher encoding accuracy indicates greater separability between human- and AI-generated responses, and thus greater dissimilarity within the linguistic feature space captured by the encoding model; lower accuracy indicates greater overlap between the two source distributions within that feature space, and thus greater similarity.

As shown in Fig. 2b, encoding models achieved high accuracy in separating human-generated responses from both GPT-4-generated responses (encoding accuracy = 0.983, 95% CI [0.958, 1.000]) and GPT-5-generated responses (encoding accuracy = 0.925, 95% CI [0.875, 0.975]). However, encoding accuracy was significantly lower for human versus GPT-5 responses (95% CI [-0.112, -0.004], $p = 0.034$), suggesting that GPT-5 responses were linguistically closer to human responses than GPT-4 responses were.

To characterize the informational structure underlying encoding differences, we next examined regression weights for individual linguistic features (Fig. 2c). Positive weights indicate that higher values of a feature increased the probability that a response was human-generated; negative weights indicate that higher values increased the probability that a response was AI-generated. As shown in Fig. 2c, GPT-4 responses were primarily characterized by greater length, fewer typos, and increased redundancy relative to human responses. In GPT-5, by contrast, encoding differences related to response length were substantially reduced, while other features, including punctuation density and typos, became more informative of the generating source.

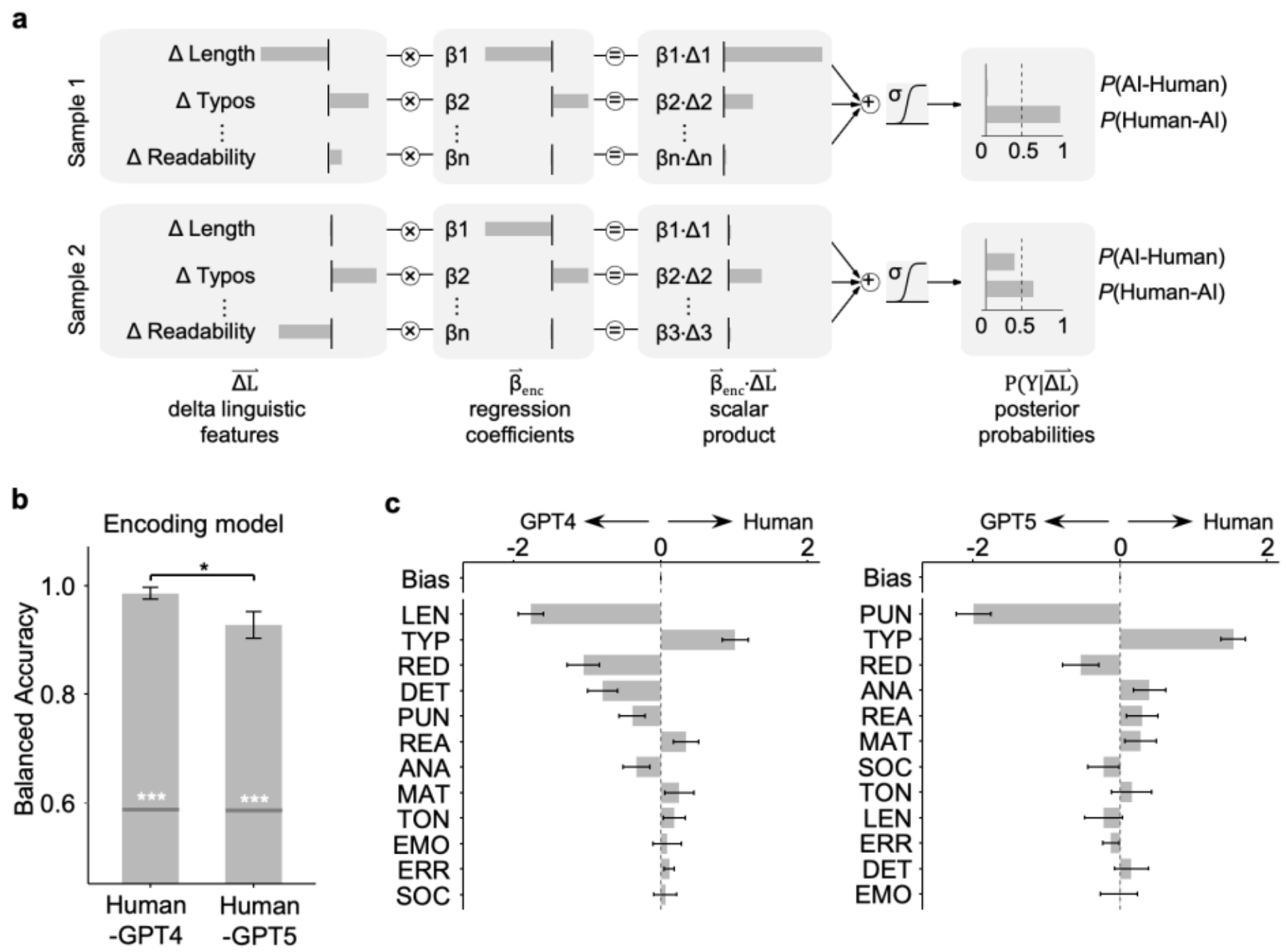

**Figure 2. Linguistic encoding of human- and AI-generated responses. a)** Schematic of the encoding model. The model uses the difference in feature values between the two presented responses (Human vs. AI) to predict the probability that the Human (or AI) response is presented in the first interval. **b**) Encoding accuracy (defined as the balanced classification accuracy of encoding models) for human vs. GPT-4 responses and human vs. GPT-5 responses. Horizontal lines within bars indicate the average value expected if there was no association between feature and label, obtained by averaging over 1,000 shuffles. Asterisks above horizontal lines denote significant deviations of observed values from the average shuffled value; asterisks above brackets indicate significant differences between conditions (*p < .05, ***p < .001). **c**) Encoding model regression weights for individual linguistic features, shown separately for GPT-4 (left) and GPT-5 (right). Positive weights indicate features for which higher values increase the probability that a response is human-generated; negative weights indicate features for which higher values increase the probability that a response is AI-generated. Error bars indicate bootstrap mean ± 1 SD. Feature abbreviations: LEN = length; TYP = typos; RED = redundancy; DET = detail; PUN = punctuation; REA = readability; ANA = analytical tone; MAT = lexical diversity; SOC = social content; TON = conversational tone; EMO = emotional content; ERR = errors.

## Readout

Having characterized the extent to which human- and AI-generated responses differed at the level of encoding, the next step was to characterize how human and AI interrogators extracted and used this source-related information to make their discrimination choices. To do so, we trained logistic regression models to predict interrogators' choices from the same set of linguistic features used in the encoding analyses. Models were trained separately for interrogators and task conditions, allowing us to estimate how different interrogators weighed and combined linguistic features when discriminating between human- and AI-generated responses in each task.

Readout models successfully predicted interrogators' choices across interrogators and tasks (Fig. S1), indicating that choices were systematically related to measurable linguistic features. Fig. 3a,b shows the contribution of individual linguistic features to discrimination choices as measured by regression weights in readout models. To assess to what extent interrogators relied on informative features, i.e., features that encoded source information, and whether they associated those features with the correct response source, we computed two indices of intersection between encoding and readout (Fig. 3c,d) [17]: $\text{overlap}_{enc,read}$, quantifying the extent to which interrogators relied on informative features irrespective of weighting direction, and $\text{alignment}_{enc,read}$, quantifying whether informative features were mapped to the correct generating source.

In the first series of studies, both human and GPT-4 interrogators relied prominently on response length (Fig. 3a), the most informative encoding feature, resulting in high overlap across interrogators and tasks (Fig. 3c; Human interrogators AI task: $overlap_{enc,read}$ = 0.862, 95% CI [0.744, 0.951], p = 0.004; Human interrogators Human task: $overlap_{enc,read}$ = 0.823, 95% CI [0.707, 0.919], p = 0.004; GPT-4 interrogators AI task: $overlap_{enc,read}$ = 0.736, 95% CI [0.575, 0.866], p = 0.004; GPT-4 interrogators Human task: $overlap_{enc,read}$ = 0.836, 95% CI [0.666, 0.953], p = 0.004). Human interrogators consistently weighted longer responses toward the AI category across both tasks, resulting in high $alignment_{enc,read}$ (Fig. 3c; AI task: $alignment_{enc,read}$ = 0.782, 95% CI [0.656, 0.884], p = 0.008; Human task: $alignment_{enc,read}$ = 0.701, 95% CI [0.548, 0.843], p = 0.008). As shown in Fig. S2c, neither $overlap_{enc,read}$ nor $alignment_{enc,read}$ varied systematically as a function of participants' self-reported use of chatbots in either task.

GPT-4 interrogators, by contrast, showed marked task-dependent (mis-)alignment (AI task: $alignment_{enc,read}$ = 0.569, 95% CI [0.386, 0.739], p = 0.008; Human task: $alignment_{enc,read}$ = -0.756, 95% CI [-0.920, -0.540], p = 0.008; difference = 1.325, 95% CI [1.038, 1.560], p = 0.004). In the AI task, GPT-4 interrogators correctly associated longer responses with AI-generated texts, whereas in the Human task they incorrectly reversed this mapping, associating longer responses with humans (Fig. 3a). This misalignment is consistent with the below-chance performance observed in the Human task (Fig. 1). Across interrogators, $overlap_{enc,read}$ did not differ in either task (AI task: p = 0.256; Human task: p = 0.827), whereas $alignment_{enc,read}$ was higher for human interrogators than for GPT-4 interrogators (AI task: p = 0.038; Human task: p = 0.004).

In the second series of studies, GPT-5 interrogators showed both greater $overlap_{enc,read}$ (AI task: p = 0.004; Human task: p = 0.004) and greater $alignment_{enc,read}$ (AI task: p = 0.004; Human task: p = 0.004) than human interrogators in both tasks. Human interrogators showed relatively weak and distributed readout weights, relying modestly on informative features such as punctuation and typographical errors (Fig. 3b; AI task: $overlap_{enc,read}$ = 0.688, 95% CI [0.535, 0.810], p = 0.026; Human task: $overlap_{enc,read}$ = 0.572, 95% CI [0.298, 0.791], p = 0.320; AI task: $alignment_{enc,read}$ = 0.619, 95% CI [0.430, 0.767], p = 0.008; Human task: $alignment_{enc,read}$ = 0.346, 95% CI [0.044, 0.595], p = 0.026). This pattern did not vary with the participants' self-reported use of chatbots in the Human task. In the AI task, however, both $overlap_{enc,read}$ and $alignment_{enc,read}$ were higher among human interrogators reporting more frequent chatbot use (Fig. S2c).

GPT-5 interrogators, by contrast, showed strong and highly aligned weighting of the features that objectively distinguished GPT-5 from human responses. In both tasks, GPT-5 interrogators consistently associated denser punctuation and greater redundancy with AI-generated responses, while associating typos with human-generated responses (Fig. 3b). This pattern resulted in both high overlap and high alignment across tasks (Fig. 3d; AI task: $overlap_{enc,read}$ = 0.933, 95% CI [0.859, 0.980], p = 0.004; Human task: $overlap_{enc,read}$ = 0.929, 95% CI [0.855, 0.979], p = 0.004; AI task: $alignment_{enc,read}$ = 0.924, 95% CI [0.840, 0.978], p = 0.008; Human task: $alignment_{enc,read}$ = 0.919, 95% CI [0.835, 0.975], p = 0.008). Consistent with the superior discrimination performance observed in Fig. 1, both $overlap_{enc,read}$ (AI task: p = 0.004; Human task: p = 0.004) and $alignment_{enc,read}$ (AI task: p = 0.004; Human task: p = 0.004) were greater for GPT-5 than for human interrogators in both tasks.

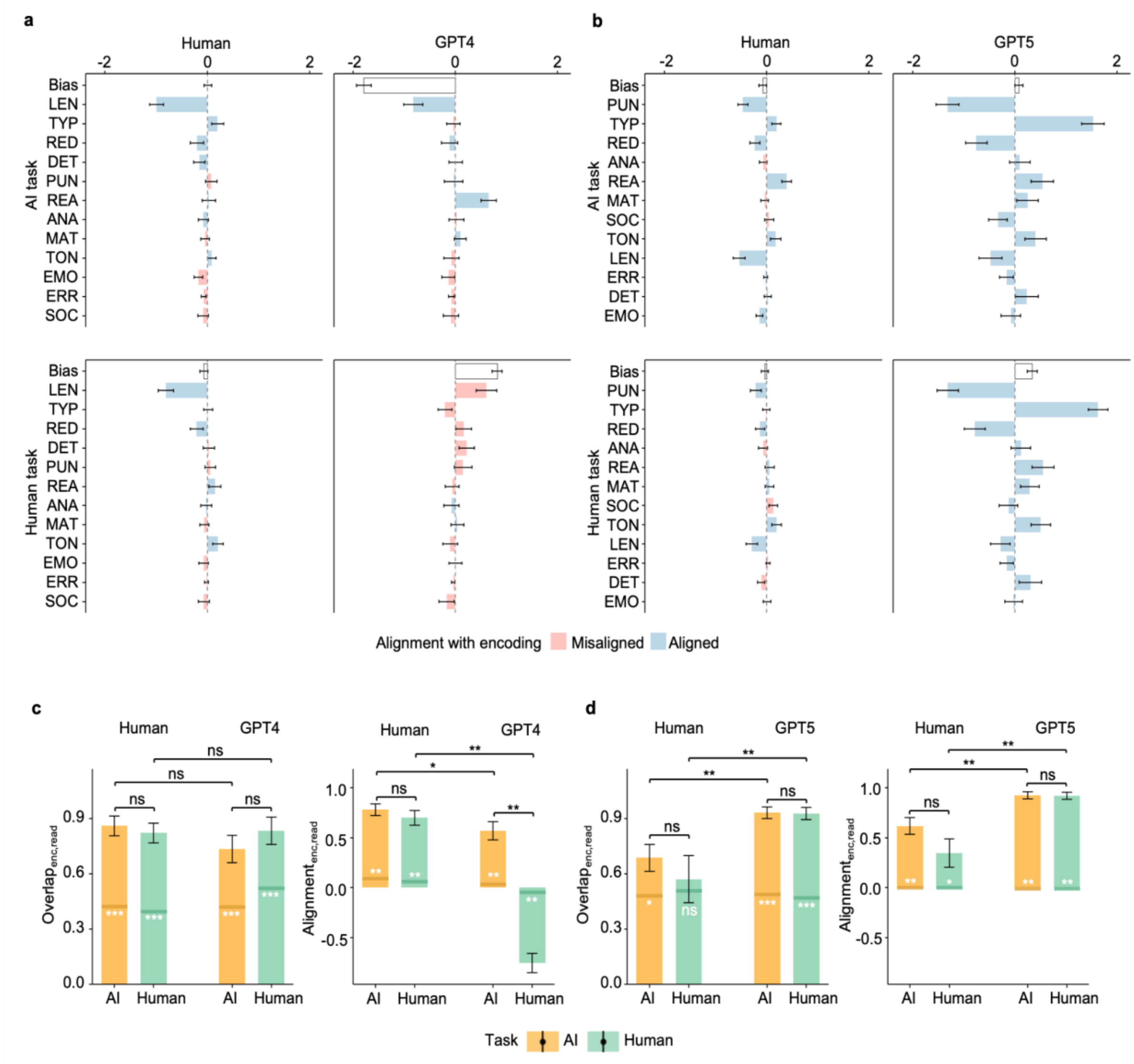

**Figure 3. Readout of source-discriminating information by human and AI interrogators. a-b)** Regression weights from readout models linking individual linguistic features to interrogators' discrimination choices, shown separately for human and AI interrogators in a) the first series of studies (GPT-4) and b) the second series of studies (GPT-5). Column labels indicate the interrogator. X-axis labels indicate the task. Positive weights indicate features for which higher values increase the probability that the response is judged as human-generated; negative weights indicate features for which higher values increase the probability that a response is judged as AI-generated. Blue bars indicate alignment with encoding and red bars indicate misalignment. Top and bottom rows correspond to the AI task and Human task, respectively. Error bars indicate the standard deviation of the bootstrap estimates. **c-d)** Quantification of $overlap_{enc,read}$ and $alignment_{enc,read}$ for human and AI interrogators in c) the first series of studies (GPT-4) and d) in the second series of studies (GPT-5). Bars represent posterior mean estimates; error bars indicate bootstrap confidence intervals. Horizontal lines within bars indicate the average value after 10,000 feature-order shuffles. Asterisks above horizontal lines denote significant deviations of observed values from the mean shuffled value; asterisks above brackets indicate significant differences between conditions (*p < .05, **p < .01, ***p < .001; ns = not significant). Feature abbreviations as in Figure 2.

## Meta-readout

Finally, we examined meta-readout, defined as the interrogators' explicit access to their readout computations. We analyzed the linguistic features that interrogators reported using to justify their judgments and compared these reported features with the actual readout weights estimated from their discrimination behavior. Using the same framework as above, we computed overlap and alignment indices between reported feature and readout weights (Fig. 4c,d). Here, $overlap_{read,meta\text{-}read}$ quantified the extent to which interrogators explicitly reported relying on the same features that actually guided their choices, irrespective of weighting direction, whereas $\text{alignment}_{\text{read,meta-read}}$ quantified whether interrogators described these features in a manner consistent with their actual readout computations.

Fig. 4a,b shows estimated meta-readout weights separately for each interrogator and task. In the first series of studies, human interrogators showed moderate $\text{overlap}_{\text{read,meta-read}}$ and $\text{alignment}_{\text{read,meta-read}}$ with their own readout structure, indicating that they were able to report readout features in a manner broadly consistent with their actual use during discrimination (Fig. 4c; AI task: $\text{overlap}_{\text{read,meta-read}}$ = 0.691, 95% CI [0.644, 0.740], $p$ = 0.004; Human task: $\text{overlap}_{\text{read,meta-read}}$ = 0.764, 95% CI [0.710, 0.812], $p$ = 0.004; AI task: $\text{alignment}_{\text{read,meta-read}}$ = 0.639, 95% CI [0.589, 0.687], $p$ = 0.008; Human task: $\text{alignment}_{\text{read,meta-read}}$ = 0.624, 95% CI [0.564, 0.682], $p$ = 0.008). This pattern did not vary with participants' self-reported use of chatbots in the AI task; in the Human task, however, $\text{overlap}_{\text{read,meta-read}}$ was higher among human interrogators reporting more frequent chatbot use (Fig. S2e).

GPT-4 interrogators showed significant $\text{overlap}_{\text{read,meta-read}}$ in the Human task ($\text{overlap}_{\text{read,meta-read}}$ = 0.621, 95% CI [0.583, 0.651], $p$ = 0.004) but not in the AI task ($\text{overlap}_{\text{read,meta-read}}$ = 0.464, 95% CI [0.413, 0.511], $p$ = 0.082). $\text{Alignment}_{\text{read,meta-read}}$ was also significant in the Human task ($\text{alignment}_{\text{read,meta-read}}$ = 0.176, 95% CI [0.122, 0.218], $p$ = 0.034) but not in the AI task ($\text{alignment}_{\text{read,meta-read}}$ = 0.095, 95% CI [0.029, 0.159], $p$ = 0.671), suggesting limited explicit access both to which features guided their decisions and to how those features were weighted during discrimination. Relative to human interrogators, GPT-4 interrogators showed both lower $\text{overlap}_{\text{read,meta-read}}$ (AI task: $p$ = 0.004; Human task: $p$ = 0.004; Fig. 4c) and lower $\text{alignment}_{\text{read,meta-read}}$ (AI task: $p$ = 0.004; Human task: $p$ = 0.004) across both tasks.

In the second series of studies, human interrogators showed significant $\text{overlap}_{\text{read,meta-read}}$ in the Human task but not in the AI task (Fig. 4d; AI task: $\text{overlap}_{\text{read,meta-read}}$ = 0.536, 95% CI [0.481, 0.594], $p$ = 0.997; Human task: $\text{overlap}_{\text{read,meta-read}}$ = 0.670, 95% CI [0.613, 0.724], $p$ = 0.034), with no difference as a function of participants' self-reported use of chatbots (Fig. S2e). GPT-5 interrogators, by contrast, showed above-chance $\text{overlap}_{\text{read,meta-read}}$ in both tasks (AI task: $\text{overlap}_{\text{read,meta-read}}$ = 0.613, 95% CI [0.594, 0.634], $p$ = 0.004; GPT-5 Human task: $\text{overlap}_{\text{read,meta-read}}$ = 0.644, 95% CI [0.628, 0.663], $p$ = 0.004), indicating that could identify several of the features that they actually read. Meta-readout overlap was higher for GPT-5 interrogators than for human interrogators in the AI task ($p$ = 0.036).

$\text{Alignment}_{\text{read,meta-read}}$, by contrast, was substantially lower for GPT-5 than for human interrogators in both tasks (AI task: $p$ = 0.004; Human task: $p$ = 0.004). Humans showed lower $\text{alignment}_{\text{read,meta-read}}$ in the AI task relative to the Human task, with $\text{alignment}_{\text{read,meta-read}}$ in the AI task being modulated by self-reported use of chatbots (Fig. S2e). GPT-5 interrogators, by contrast, showed similarly weak correspondence between reported and actual feature weighting in both the AI task and the Human task (AI task: $\text{alignment}_{\text{read,meta-read}}$ = 0.297, 95% CI [0.273, 0.323], $p$ = 0.008; Human task: $\text{alignment}_{\text{read,meta-read}}$ = 0.296, 95% CI [0.273, 0.320], $p$ = 0.008), with remarkably similar misalignment patterns across the two task conditions. In both tasks, for example, GPT-5 reported treating greater analytical tone, higher readability, and greater detail as cues to AI-generated text, whereas its actual readout weights associated higher values of these features with human-generated text. This suggests a dissociation between $\text{overlap}_{\text{read,meta-read}}$ and $\text{alignment}_{\text{read,meta-read}}$ such that GPT-5 interrogators correctly reported which features they used to distinguish human- from AI-generated responses, but not how they used them.

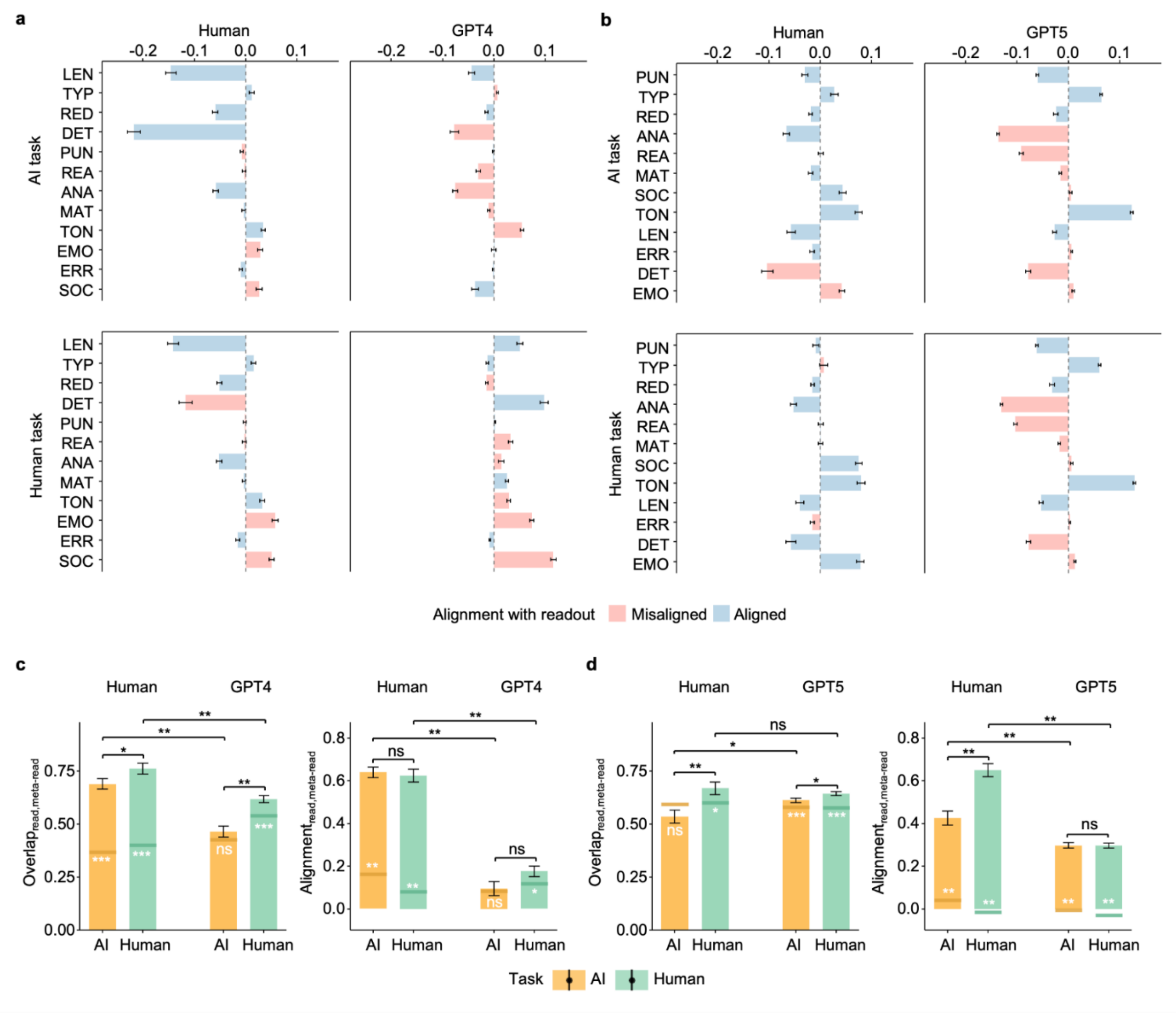

**Figure 4. Meta-readout. a–b)** Feature weights derived from interrogators' verbal explanations, shown separately for human and AI interrogators in a) the first series of studies (GPT-4) and b) the second series of studies (GPT-5). Column labels indicate the interrogator. X-axis labels indicate the task. For each trial, explanation vectors were L1-normalized and averaged within each interrogator-by-task condition to yield a condition-level reasoning vector. Positive weights indicate features for which higher values were reported as evidence for human authorship; negative weights indicate features for which higher values were reported as evidence for AI authorship. Blue bars indicate that the reported direction of a feature was aligned with the corresponding readout weight; red bars indicate misalignment. Upper and lower rows correspond to the AI task and Human task, respectively. Error bars indicate the standard deviation of the bootstrap estimates. **c–d)** Quantification of $\text{overlap}_{\text{read,meta-read}}$ and $\text{alignment}_{\text{read,meta-read}}$ for human and AI interrogators in c) the first series of studies (GPT-4) and d) the second series of studies (GPT-5). Overlap quantifies whether interrogators explicitly mentioned the features that predicted their choices, irrespective of direction. Alignment quantifies whether the reported direction of those features matched the direction in which they were used in the readout model. Bars represent posterior mean estimates; error bars indicate bootstrap confidence intervals. Horizontal lines within bars indicate the average value after 10,000 feature-order shuffles. Asterisks above horizontal lines denote significant deviations from the mean shuffled value; asterisks above brackets indicate significant differences between conditions (*p < .05, **p < .01, ***p < .001; ns = not significant). Feature abbreviations as in Figure 2.

## Discussion

Across two generations of large language models, we show that behavioral distinguishability between human- and AI-generated responses decomposes into dissociable contributions from encoding, readout, and meta-readout.

At the level of encoding, accuracy declined across model generations, indicating that, although AI-

generated and human responses remained statistically distinguishable, their similarity within the linguistic feature space captured by the encoding model increased from GPT-4 to GPT-5. This convergence cannot be attributed to an explicit trial-level instruction to imitate human responses. Rather, it likely emerges as a byproduct of broader improvements in generative capability of LLMs, analogous to the gains across model generations observed in theory-of-mind[18]. The informational structure distinguishing the two sources also shifted across model generations. Specifically, whereas GPT-4 responses were primarily characterized by greater length relative to human responses, GPT-5 responses no longer differed in length and were distinguished instead by denser punctuation and fewer typographical errors.

At the level of readout, this shift had asymmetric consequences for human and AI interrogators. As GPT-5 responses became more similar to human responses, discriminating between the two sources became intrinsically more difficult for humans. This is consistent with the general principle that behavioral discriminability declines as overlap between stimulus distributions increases [19, 20]. Examination of the readout weights revealed that human interrogators made weak use of the features that most strongly distinguished GPT-5 from human responses, including punctuation density and typographical errors, while continuing to assign substantial weight to response length, a feature diagnostic of GPT-4 responses but no longer reliably distinguishing GPT-5 from human responses. This mismatch raises the possibility that human interrogators relied on heuristics calibrated to the distributional differences characteristic of earlier generations of LLMs. Under this interpretation, the decline in human discrimination performance across model generations would reflect not only weaker encoding differences, but also a failure to update readout strategies as those differences change. Analysis of chatbot-use frequency supports this interpretation: participants reporting greater chatbot use did not show the same legacy weighting of response length in the AI task, and instead correctly weighted currently diagnostic features, including punctuation density and typographical errors, consistent with the idea that greater exposure to chatbots facilitates the recalibration of readout strategies.

GPT-5 interrogators, by contrast, achieved high and symmetric readout overlap and alignment across task conditions. Despite the increased difficulty of the task, GPT-5 interrogators showed enhanced, not diminished, extraction and processing of source-discriminating information. Improved generation and improved readout thus co-occurred within GPT-5. This pattern, which runs directly counter to the decline observed in human interrogators, has an instructive parallel with the acquisition of expertise in humans. Across domains as different as sport, music, chess, and medicine, human experts not only perform better than novices, but develop an enhanced perceptual sensitivity [21]. It is tempting to speculate that GPT-5 results may reflect an analogous form of machine expertise: as generative representations become more deeply aligned with the statistical structure of human language, readout sensitivity to the subtle features distinguishing AI- from human-generated text simultaneously scales. This raises the possibility that newer LLMs are becoming expert readers of the human textual space, capable of exploiting residual differences that escape non-expert observation.

At the level of meta-readout, however, the GPT-5 advantage relative to humans reversed. Despite achieving near-ceiling readout alignment, GPT-5 showed substantially lower meta-readout alignment than human interrogators. Taken together, these findings suggest a double dissociation between readout and meta-readout alignment across interrogators: human interrogators show modest readout alignment but comparatively higher meta-readout alignment; GPT-5 interrogators show high readout alignment but comparatively lower meta-readout alignment.

This dissociation between what GPT-5 interrogators *do* and *what they say they did* parallels recent findings showing that reasoning models often fail to disclose the use of information that demonstrably influences their responses [22] and, more broadly, that LLM self-explanations may be plausible without being faithful to the computations that actually drive their outputs [23]. This raises the more fundamental concern that, in our data, meta-readout may reflect an entirely post-hoc rationalization decoupled from the readout. If this were the case, however, we would expect both meta-readout overlap and alignment to be at chance. This was not the case. GPT-5 interrogators reliably reported many of the features that they used, yielding meta-readout overlap well above chance. Meta-readout alignment, while lower than in human interrogators, was also above chance. This pattern argues against meta-readout as a wholly confabulated account of readout and suggests instead a failure of alignment confined to a specific subset of features, including analytic tone, readability, and detail. Why these specific dimensions break down, whether they share similar computational properties, and whether a similar misalignment pattern is common to different model families, remain compelling questions for future research.

At the level of human-AI interaction, a localized breakdown of alignment may introduce insidious meta-cognitive traps [14] and contribute to failures of coordination [24]. Users may infer that the model's reasoning is aligned with their own, even when it maps some of those features to the opposite conclusion. This illusion of alignment may be especially consequential as people increasingly consult LLMs when making decisions or delegate decisions to LLMs altogether [25].

The Turing Test has been described as a test "not for AI to pass, but for humans to fail" [3]. Here, we make two moves. First, we invert the test, placing humans and AI in symmetrical roles as interrogators. Second, we decompose failure into distinct contributions from encoding, readout, and meta-readout. This reveals how failure at one level can coexist with success at another, moving beyond the binary logic of pass or fail toward a mechanistic account of how human and artificial systems encode information about generating sources, how they read this information out, and how much access they have to the underlying readout computations. These metrics may prove useful beyond the Turing Test setting, as they provide a principled method for 'breaking' information processing and opening the black box of LLMs.

## Methods

We conducted two series of studies. In the first, human- and GPT-4-generated responses were evaluated by both human and GPT-4 interrogators; in the second, human- and GPT-5-generated responses were evaluated by both human and GPT-5 interrogators. The research protocol was approved by the local ethics committee (Ethik-Kommission der Ärztekammer Hamburg) and complied with the principles of the revised Helsinki Declaration [26].

### Experimental models and subject details

**Human interrogators.** Human interrogators were recruited through the online recruitment platform Prolific. Participants were required to be native English speakers between 18 and 70 years of age, with no reported history of psychiatric conditions and no reported history of dyslexia. Participants who had previously taken part in [27] and participants who had completed the Strange Stories test used to generate stimuli to be used for the GPT-5 comparison (see Stimuli), were excluded from serving as interrogators. The sample size was determined by the number of stimulus pairs and by the fully counterbalanced design. Participants were instructed to complete the experiment on a laptop or desktop computer; those detected as using a mobile browser were excluded and replaced. The final sample included 1,904 participants for the GPT-4 comparison and 1,920 participants for the GPT-5 comparison, with each participant contributing a single judgment.

**AI interrogators.** We tested two generations of OpenAI GPT models as AI interrogators. In the GPT-4 comparison, the primary AI interrogator was GPT-4 (*'gpt-4-0613'*), the same model generation that produced the AI answers in the stimulus set in (Strachan et al., 2024). Data from GPT-4 interrogators were collected in November 2024. In the GPT-5 comparison, the primary AI interrogator was GPT-5 (*'gpt-5-2025-08-07'*). Data from GPT-5 interrogators were collected in September 2025. LLM interrogators were queried through the OpenAI API using a custom script. Sampling parameters were held constant across model interrogators (temperature = 0.8, top-p = 0.9).

### Stimuli

Stimuli were drawn from the Strange Stories test [15], a theory-of-mind paradigm that uses short social scenarios involving situations such as deception, irony, persuasion, misunderstanding, and mistaken identity. Each item ends with a question that requires an open-ended description of a character's mental state or social action. We selected this test because it elicits extended textual answers, making it suitable for quantifying linguistic features that may differentiate human- and AI-generated responses. In both comparisons, we retained only responses coded as correct under the original test scoring criteria (with a score of 2), ensuring that interrogators could not solve the discrimination task by identifying incorrect mental-state reasoning.

For the GPT-4 comparison, we used the dataset reported by [27], which included 400 human responses from 50 participants completing 8 Strange Stories items and 120 GPT-4 responses collected across 15 GPT-4 sessions. One GPT-4 response was excluded because it did not meet the correctness criterion, leaving 119 correct GPT-4 responses. Human correct responses (n = 295) were randomly down sampled within item and paired with the correct GPT-4 responses, yielding 119 human–GPT-4 response pairs.

For the GPT-5 comparison, we collected a new dataset using the same protocol, comprising 424 human responses from 53 participants and 120 GPT-5 responses to the same 8 Strange Stories items. Nine human participants were excluded because they had previously completed the task. All 120 GPT-5 responses met the correctness criterion. Human correct responses (n = 267) were again randomly down sampled within item and paired with the correct GPT-5 responses, yielding 120 human–GPT-5 response pairs.

**Turing and Inverse Turing Test protocol**

Each participant (human or AI) completed a single trial consisting of three sequential steps: an authorship decision, a confidence rating, and a written explanation. At the beginning of each trial, interrogators were told that they would read a short story involving human interactions, followed by two responses to the same question, one produced by a human and one by AI. Following this preamble, the original text of the Strange Stories item was presented. The two answers were shown in one of two possible orders, human–AI or AI–human. Depending on the task condition, interrogators were instructed to identify either the response produced by the human (Human task) or the response produced by AI (AI task). Thus, the decision was always a two-interval forced-choice judgment: which of the two responses, first or second, corresponded to the instructed source. After making their decision, interrogators rated their confidence on a scale from 1, very low, to 100, very high. Finally, they provided a written explanation of how they made their choice: "*Describe how you made your decision. Please, provide specifics of anything in the text that stood out to you as particularly human or AI-like. Be clear and legible in your response, as it will be analyzed and coded by human researchers.*"

**Procedure**

Human interrogators completed the experiment online in SoSci. The decision, confidence rating, and explanation were presented on consecutive pages. LLM interrogators were tested through the OpenAI API. The task sequence was matched to the human experiment and implemented as chained prompts: first the generating source decision (including both human- and AI-generated responses within the same prompt), then the confidence rating, and finally the written explanation. This ensured that confidence and explanation were conditioned on the preceding decision. If a model failed to provide a valid forced-choice response, the trial was rerun until the corresponding design cell was complete.

Answer order and task instruction were fully counterbalanced. Each stimulus pair was presented in both possible answer orders (human–AI, AI–human) and under both task instructions (identify human, identify AI). Each combination of response pair, answer order, and instruction was completed by four human interrogators and four LLM interrogator runs. Thus, in the GPT-4 comparison, 119 response pairs yielded 1,904 observations from human interrogators and 1,904 from GPT-4 interrogators. In the GPT-5 comparison, 120 response pairs yielded 1,920 observations from human interrogators and 1,920 from GPT-5 interrogators.

**Quantification and data analysis**

**Signal detection theory.** Responses were first analyzed using signal detection theory (SDT), which separates discrimination sensitivity from response bias (Green & Swets, 1966). Because the task was a two-interval forced-choice design, responses were recorded separately for each task by treating the instructed target in the first position as the signal. In the AI task, the AI answer appearing first was treated as the signal; in the Human task, the human answer appearing first was treated as the signal. A response of "first" was coded as a hit when the target was in the first position and as a false alarm when the target was in the second position.

We computed sensitivity (d') and criterion (c). Sensitivity was computed as the difference between the z-transformed hit and false-alarm rates and indexed the ability to discriminate the target author independently of response position. Criterion was computed as the negative half-sum of the z-transformed hit and false-alarm rates and indexes the tendency to select one position over the other. Extreme hit and false-alarm rates were corrected using the Hautus adjustment [28].

SDT indices were computed separately for each stimulus pair, interrogator type, and task.

**Identifying and quantifying diagnostic features.** To quantify how linguistic properties contributed to source encoding, readout, and meta-readout, we constructed a common set of 12 features (see below) capturing the cues that interrogators themselves reported using. Candidate features were first identified through preliminary coding of explanations provided by human and AI interrogators. Analytic tone, punctuation, conversational language, emotion references, and social-process language were quantified using LIWC-22, a validated and widely used toolbox for quantifying psychological dimensions in language use [29]. LIWC-22 part-of-speech measures were also used in the computation of response detail. Response length, lexical diversity, readability, contractions, and typographical errors were computed automatically in Python using standard text-processing packages. Features that could not be captured reliably using automated procedures, specifically redundancy and factual errors, were coded manually. The operational definition of each feature is provided below.

**Length (LEN).** Length was parameterized by computing the word count of each response using the Python `.split()` command.

**Detail (DET).** We considered the level or amount of detail of a response as referring to the amount of new information that was expressed in the text. Words that convey additional information are those that narrow down the space of possibilities, conveying precise information about people, objects, and actions within a scene. In order to quantify the amount of detail in a way that was dissociated from the length of the responses, we filtered each response to keep only the unique tokens and used the LIWC-22 toolbox to get normalized frequencies of different parts of speech (POS; verbs, adverbs, adjectives). LIWC-22 does not tag nouns in its POS-tagging, but we computed the frequency of nouns by calculating the delta between the normalized frequency of dictionary-identified words and those that were POS-tagged as function words, which gave us the normalized frequency of nouns. The level of detail was then computed as the sum of the proportion of unique tokens in the text that were information-carrying (nouns, verbs, adverbs, or adjectives).

**Typos (TYP).** Typographical errors such as grammatical errors, spelling errors, or punctuation errors were computed using the `language_tool_python` package in Python. This tool, which acts as an automated US-English spelling and grammar checker, identified typos in the response texts. The results of this were then cross-checked manually by a native English speaker. The frequency of typos was then normalized to the length of the response.

**Redundancy (RED).** This feature was defined as the tendency of a response to repeat information unnecessarily or to include additional but redundant details of the story in the response (such as repeating or rephrasing the question before answering). Given that this repetition was often also associated with rephrasing, automated approaches to detect repeated n-grams were not suitable. Redundancy was quantified using manual labelling, where an experimenter highlighted the parts of the answer that repeated or rephrased previous points in a redundant way.

**Analytic tone (ANA).** This feature was quantified using the LIWC-22 toolbox to identify the normalized frequency of terms associated with analytical or logical thinking.

**Punctuation (PUN).** This feature was quantified using the LIWC-22 toolbox to identify the normalized frequency of punctuation marks within the text.

**Lexical diversity (MAT).** Diversity of the vocabulary used in responses was computed using the moving-adjustment type-token ratio (MATTR). This is a measure used to estimate the lexical diversity of a text that is independent of the length of the text. It is computed by taking the average of the type-token ratios (TTRs; the ratio of the number of unique tokens to the total number of tokens) of a moving window.

**Readability (REA).** The complexity of the sentence structure was calculated using the Flesch-Kincaid readability score, which computes the 'readability' of a text using the number of words per sentence and the number of syllables per word. Lower values of readability therefore reflect more complex sentence structures.

**Conversational tone (TON).** This feature was quantified using the LIWC-22 toolbox to identify the normalized frequency of conversational terms in the text. In addition to this, the number of contractions (e.g. "can't",'"won't") was computed using Python and normalized to the word count. The parameter of tone was the sum of these normalized values.

**Emotion references (EMO).** This feature was quantified using the LIWC-22 toolbox to identify the normalized frequency of terms associated with emotions.

**Social awareness (SOC).** This feature was quantified using the LIWC-22 toolbox to identify the normalized frequency of terms associated with social processes.

**Errors (ERR).** Factual errors (reporting details of the story incorrectly such as mixing up characters' names while being correct about the rest of the mentalistic reasoning) were rare in the responses and were coded manually by an experimenter.

**Encoding and readout models**

For each response pair, we computed a single-trial linguistic feature-difference vector, defined as the signed difference in linguistic feature values between the human response and the AI response. To quantify how much information about response source was present in each pair, we then trained logistic regression encoding models to predict the probability that in a given trial the human-generated response appeared in the first interval as a function of the linguistic feature-difference vector of that trial. Regression weights in encoding models provide a measure of the contribution of linguistic features to source information encoding. The encoding vector, with components equal to the regression weights of the encoding model, specifies the weighted combination of features that best predicts the source of the response.

Similarly to information encoding, to quantify information readout, we trained logistic regression models to predict the interrogator's discrimination choice in each trial as a function of the linguistic feature-difference vector of that trial. Separate readout models were fitted for each interrogator type, task, and stimulus set. Regression weights in readout models provide a measure of the contribution of linguistic differences to source information readout. The readout vector, with components equal to the regression weights of the readout model, specifies the weighted combination of features that best predicts the choice of the interrogator.

**Model construction.** All encoding and readout models were fitted as ridge-regularized logistic regressions using the *glmnet* package in R (Friedman et al., 2010). Predictors were z-scored before fitting, and performance was evaluated using balanced accuracy to avoid dependence on any class imbalance. Models used regularized logistic regression implemented with L2 (ridge) regularization to handle potential multicollinearity among language features while retaining all features and avoiding coefficients being reduced to zero. The strength of regularization was controlled by the hyperparameter λ. To select this hyperparameter, we evaluated 100 candidate λ values sampled on a logarithmic grid from $10^{-3}$ to $10^{3}$. We selected the optimal value of the regularization strength λ as the one maximizing the cross-validated balanced accuracy combined across the primary readout models. The selected value was λ =1.23285e-2. The same λ was used for all encoding and readout models.

**Overlap and alignment between encoding and readout.** To quantify how well interrogators read the encoded source information, we computed two indices of intersection between encoding and readout: $overlap_{enc,read}$ and $alignment_{enc,read}$.

Encoding-reaodut overlap, $overlap_{enc,read}$, measured the extent to which interrogators relied on linguistic features that encoded source information, irrespective of the direction in which those features were weighted. It was computed as the cosine similarity between the elementwise absolute value of the encoding and readout vectors. This index ranges from 0 to 1, with higher values indicating that interrogators assigned greater weight to the same features that distinguished human- from AI-generated responses.

Encoding-readout alignment, $alignment_{enc,read}$, measured whether interrogators weighted source-informative linguistic features in the direction specified by the encoding model. It was computed as the cosine similarity between the signed encoding and readout vectors. This index ranges from -1 to 1 and quantifies whether the two vectors weigh the same features in the same direction. Positive values indicate that interrogators used source-informative features in a direction consistent with the encoding structure; negative values indicate that they reversed this mapping; values close to zero indicate weak directional correspondence between encoding and readout.

**Coding of free-text explanations.** To identify which features interrogators explicitly reported as informing their decisions, we applied an LLM-assisted coding procedure to each free-text explanation. Explanations were coded for two aspects: feature mention, indicating whether a given feature was described as relevant to the choice, and feature direction, indicating whether higher values of that feature were interpreted as evidence for AI or human authorship.
Coding was performed using GPT-4o within a custom Python pipeline. For each explanation, the model evaluated the predefined linguistic features one at a time and assigned a directional code to each feature: 1 if higher values of the feature were interpreted as evidence for AI authorship, −1 if higher values were interpreted as evidence for human authorship, and 0 if the explanation provided no evidence that the feature informed the decision. This coding step was repeated 10 times with a temperature of 1.0. The resulting codes were then passed to a second model, run at a temperature of 0.4, which aggregated the codes of the first agent and selected a final code for each feature. When agreement across the repeated classifications fell below a predefined threshold of 0.85, the model was reinvoked at temperature 0 to produce a deterministic final code. The output and reasoning of the second model was exported alongside the data.

The coding prompt included detailed definitions of each linguistic feature, together with experimenter-generated examples illustrating typical and limiting cases. To validate the LLM-assisted coding, two experimenters independently coded a subset of 379 explanations, corresponding to 10% of the full dataset. Because some features in the dataset were rare, agreement between human and GPT4o coding was quantified using the Gwet's Agreement Coefficient 1 (AC1) [30], which is less sensitive than Cohen's kappa to prevalence imbalance. Two AC1 scores were computed: mention AC1, reflecting agreement on whether a feature was mentioned in an explanation, and category AC1, reflecting agreement on the direction assigned to the feature when both coders mentioned it. The human-coded subset was used to validate and iteratively refine the coding instructions until agreement between human and LLM coding was high (mention AC1=0.833; category AC1 = 0.967). Once this threshold was reached, the same instructions were used to code the full dataset. The agreement between the 10% subset and the same subset from the full dataset coding was also high (mention AC1 = 0.958; category AC1 = 0.982).

**Overlap and alignment between readout and meta-readout.** The coded explanations produced one value per linguistic feature for each trial. A feature received a directional code when the explanation indicated that the feature was used to support human or AI attribution. Features that were not mentioned were treated as missing at the coding stage. Features that were mentioned without a clear direction were assigned a value of 0. Before analysis, codes were transformed to match the sign convention of the readout models, such that positive values indicated evidence for human-generated response and negative values indicated evidence for an AI-generated response. For each explanation, the resulting feature vector was L1-normalized by dividing each non-zero feature value by the sum of the absolute values in that explanation. This converted each explanation into a proportional directional feature vector, so that explanations mentioning different numbers of features could be compared on the same scale. Missing feature mentions were treated as 0 when constructing these vectors. Directionless mentions coded as 0 were therefore also assigned zero directional weight; these cases were extremely rare and accounted for only 0.03% of coded feature entries.

The L1-normalized explanation vectors were then averaged within each interrogator-type-by-task condition to obtain an average meta-readout explanation vector (hereinafter, meta-readout vector). The meta-readout vector describes, for each interrogator type and task, which linguistic features were explicitly reported as informing source-discrimination choices and in which direction those features were reported to support the decision. Positive values indicate that a feature was, on average, described as evidence for a human-generated response; negative values indicate that it was described as evidence for an AI-generated response.

To quantify the correspondence between explicit explanations and implicit decision strategies, each meta-readout vector was compared with the corresponding readout vector using overlap and alignment metrics similar to those used for the encoding–readout analysis.

Readout-meta-readout overlap, $\text{overlap}_{\text{read,meta-read}}$, quantified whether interrogators explicitly reported the same features that predicted their choices, irrespective of direction. This index takes values between 0 and 1 and was computed as the cosine similarity between the elementwise absolute values of the readout and meta-readout vectors.

Readout-meta-readout alignment, $\text{alignment}_{\text{read,meta-read}}$, quantified whether the reported direction of those features matched the direction in which they were used by the readout model. This index takes values between -1 and 1 and was computed as the cosine similarity between the signed readout and meta-readout vectors.

**Use of chatbots.** Human interrogators reported how frequently they used chatbots on a six-point scale: 1, never; 2, a few times a year; 3, a few times a month; 4, a few times a week; 5, daily; and 6, many times a day. To evaluate the influence of use on readout and meta-readout, they were divided into low-frequency users (ratings 1–3) and high-frequency users (ratings 4–6). Signal-detection, readout, and meta-readout analyses were then repeated separately for the two groups using the same modelling and bootstrap procedures as in the primary analyses.

**Bootstrap statistics.** For SDT indices, encoding and readout model accuracies, regression coefficients, overlap indices, and alignment indices, means, standard errors, and confidence intervals were estimated using nonparametric bootstrap resampling [31] performed at the stimulus-pair level. For each sampled pair, all available trials corresponding to that pair were included in the bootstrapped dataset, and the relevant quantities were recomputed. Estimates were based on 1,000 bootstrap resamples.

Statistical tests against reference values, as well as pairwise comparisons between conditions, were performed nonparametrically from the corresponding bootstrap distributions. For SDT quantities, the reference value was zero. For encoding, readout, and meta-readout quantities, the reference value was estimated using permutation procedures, as described below. Empirical p-values were computed from the bootstrap samples using two-sided tests with an add-one correction for finite resampling, taking zero or the average shuffled value as the reference value as appropriate [32], and were then corrected for multiple comparisons using the Holm-Bonferroni procedure.

**Permutation-based computation of encoding, readout, and meta-readout values expected with random associations.** Balanced accuracy and overlap are non-negative and thus have non-zero expected values in case of random associations between variables. To avoid inflating evidence for effects on this basis, we conservatively referenced all encoding, readout, and meta-readout quantities against permutation-derived null distributions rather than zero. For encoding and readout accuracy, we generated label-shuffled data by randomly permuting the relevant labels relative to the linguistic feature vectors and then reran the modelling procedure using the same fixed $\lambda$. This yielded a set of balanced accuracy values expected when linguistic features carried no information about the target label. For overlap and alignment indices, we generated feature-shuffled values by randomly permuting the feature order of one vector relative to the other. Unless otherwise stated, mean shuffled values were estimated using 10,000 permutations.

**Data and code availability**

The data and code used for all analysis in the main manuscript and Supplementary Information will be made available upon publication.

**Acknowledgements**

This work is supported by the European Commission through Project ASTOUND (101071191—HORIZON-EIC-2021-PATHFINDERCHALLENGES-01 to F.M., A.R. and C.B.). The funders had no role in the study design, data collection and analysis, decision to publish or preparation of the manuscript.

## Supplementary Information

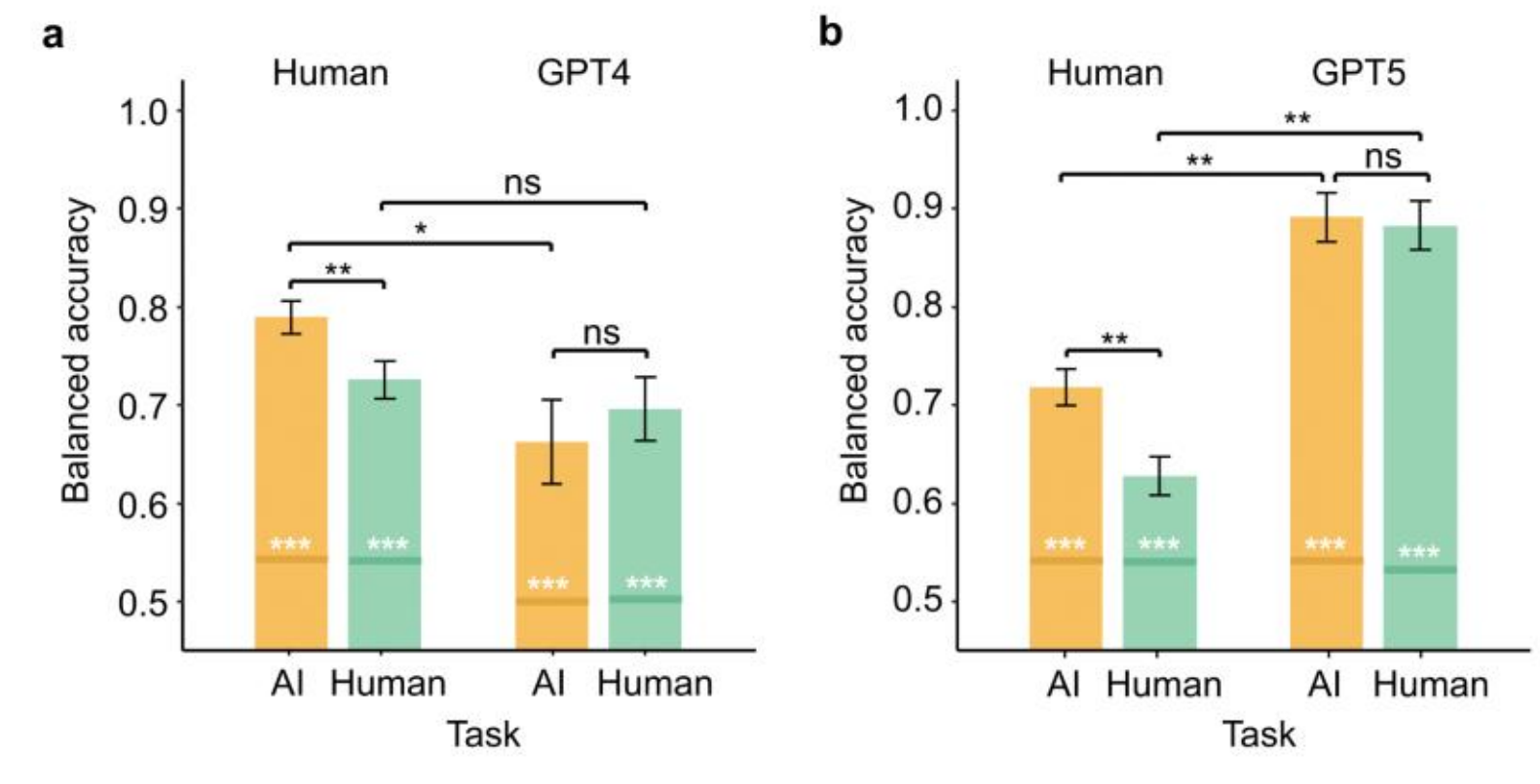

**Figure S1. Readout models successfully predicted interrogators' choices across interrogators and tasks.** Balanced classification accuracy of readout models predicting interrogators' choices from linguistic feature differences, shown separately for **a)** the human–GPT-4 comparison and **b)** the human–GPT-5 comparison. Column labels indicate the interrogator, and the x-axis indicates the task. For comparison, horizontal lines within bars indicate the average value expected if there was no association between linguistic features and interrogator choices, obtained by averaging over 1,000 shuffles. Bars show bootstrap means and error bars show ±1 SD across 1,000 stimulus-pair bootstrap resamples. Asterisks above horizontal lines denote significant deviations of observed values from the average shuffled value; asterisks above brackets indicate significant differences between conditions (*p < .05, **p < .01, ***p < .001; ns = not significant).

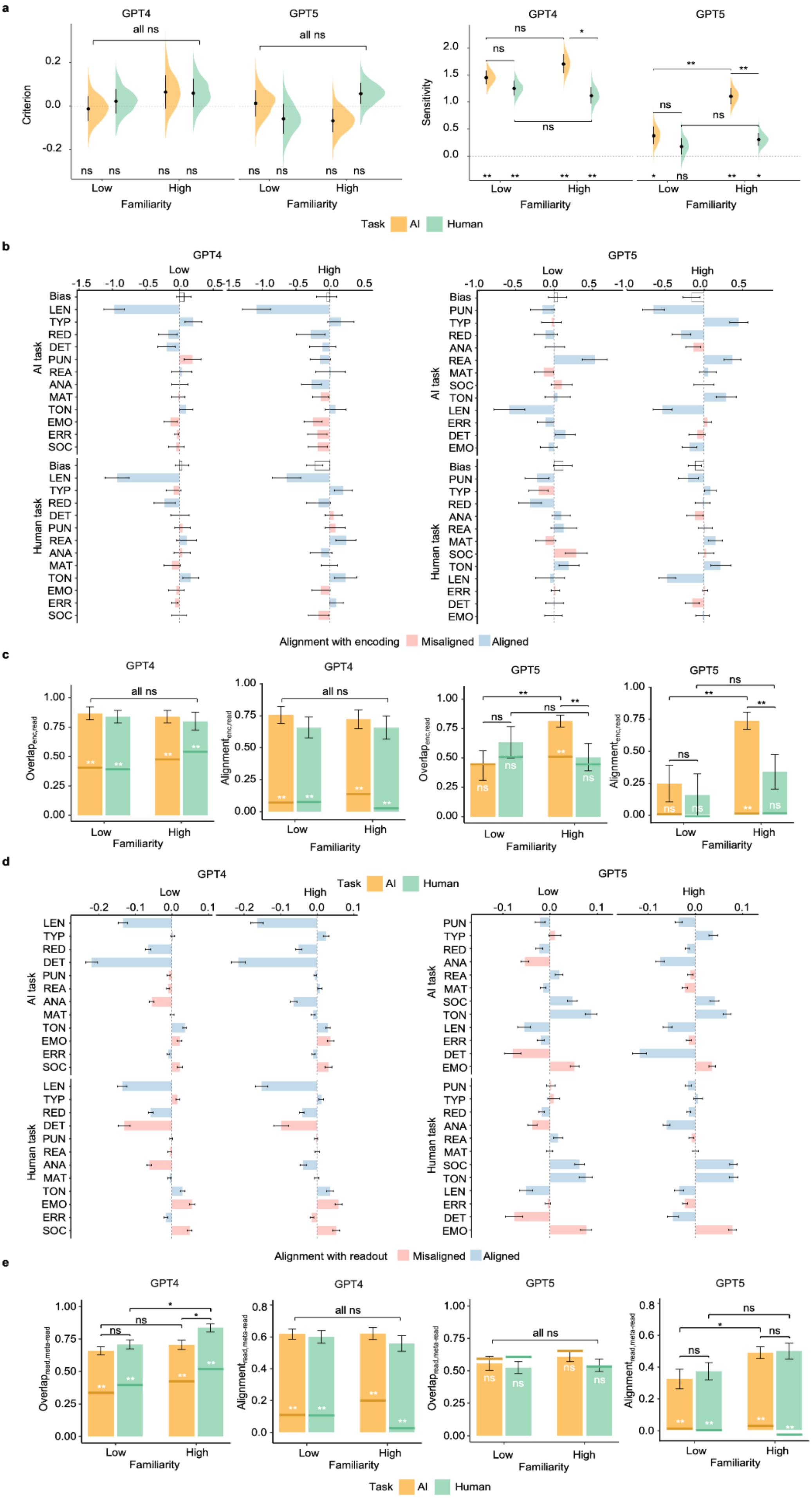
a
GPT4
GPT5
Criterion
Sensitivity
all ns
Low
High
Familiarity
Task AI Human
b
Alignment with encoding Misaligned Aligned
AI task
Human task
Bias
LEN
TYP
RED
DET
PUN
REA
ANA
MAT
TON
EMO
ERR
SOC
c
$Overlap_{enc,read}$
$Alignment_{enc,read}$
d
Alignment with readout Misaligned Aligned
e
$Overlap_{read,meta-read}$
$Alignment_{read,meta-read}$
Task AI Human

**Figure S2. Human interrogators' discrimination, readout, and meta-readout as a function of familiarity with LLMs.** Human interrogators were split into low-familiarity (ratings 1–3) and high-familiarity (ratings 4–6) groups according to their self-reported frequency of LLM use. Results are shown separately for the first series of studies (human vs. GPT-4) and the second series of studies (human vs. GPT-5). **a)** Signal Detection Theory estimates of criterion and sensitivity. **b)** Readout weights estimated separately for low- and high-familiarity human interrogators in the first series of studies. **c)** $\text{Overlap}_{\text{enc,read}}$ and $\text{alignment}_{\text{enc,read}}$. **d)** Meta-readout weights derived from human interrogators' verbal explanations. **e)** $\text{Overlap}_{\text{read,meta-read}}$ and $\text{alignment}_{\text{read,meta-read}}$. Error bars indicate ±1 SD across bootstrap estimates. Horizontal lines within bars indicate the mean shuffled value. Asterisks below or within bars indicate significant deviations from zero or from the shuffled value (*p < .05, **p < .01; ns = not significant).